# Surprise-Gated Memory for Live Speaker Diarization

An online diarizer with intra-meeting episodic memory and sleep consolidation on a frozen backbone

**Louis Mouchon** **Alexis Lainault**

Independent Research

September 26, 2026

## Abstract

We transpose the Surprise-Gated Memory recipe, a frozen encoder plus a calibrated novelty signal that gates a one-shot episodic store consolidated by periodic replay, from continual visual learning to live speaker diarization. The moment-to-moment question a streaming diarizer must answer, whether an arriving voice belongs to an already-heard speaker or a new one, is a known-versus-novel decision, exactly the decision the surprise gate is built for. On a frozen fine-tuned DiariZen backbone, our online system SGM-Diar recovers 4.8 of the 9.2 diarization-error points separating a naive online baseline from the offline oracle on AMI-SDM (52%) and 1.9 of 2.3 points on VoxConverse (82%), in the end-of-meeting revisable mode, with calibration frozen on development data and transferred cross-domain without adjustment. These are single-seed runs on three AMI meetings and five VoxConverse sessions, so the ordering of conditions rather than the absolute rates is the result. Beyond error rate, we measure the gate's behaviour directly and find a coherent memory-like profile: one-shot speaker recall at 0.857, a novelty response that habituates as speakers become familiar, re-identification that is invariant to the length of intervening silence, and a false-novelty rate of 0.001. Two ablations reproduce, now on audio, the central prediction of the visual SGM work: a sliding replay window is worse than no replay at all. The memory is strictly intra-meeting and is wiped at the end, and no gradient touches the run-time system. We report SGM-Diar as a proof-of-concept and state its limitations explicitly.

## 1 Introduction

A streaming diarizer must decide, for every fragment of speech, whether the voice belongs to a speaker already heard in the meeting or to a new one, and it must decide before hearing what comes next. An offline system escapes this: it reads the whole meeting and clusters at leisure. The gap between the two is the cost of committing in real time, and closing it is the practical problem of live diarization.

This note follows a single thread from earlier work on Surprise-Gated Memory (SGM), where a small predictor over the latent space of a frozen encoder yields a novelty signal that gates a non-parametric episodic store, consolidated offline by full replay [1]. That thread was developed for continual visual learning and for metacognition in a vision-language model. The observation here is that the known-versus-novel decision at the heart of online diarization is the same decision the surprise gate already makes, so the recipe transfers directly to audio: a frozen speaker-embedding backbone replaces the frozen vision encoder, speaker prototypes replace class prototypes, and the periodic sleep that consolidates a continual stream becomes the consolidation that stitches local speaker decisions into global identities within a meeting.

Nothing in the run-time system is trained. The backbone stays frozen, which by construction rules out representation-level forgetting, and all adaptation happens in a non-parametric memory held in RAM for the duration of a single meeting. This matches a hard product constraint: speaker memory must not persist across meetings, and it must be destroyable on demand.

We make three contributions. First, we show that the SGM surprise gate calibrated once on development data in the backbone's own scoring geometry, paired with full-buffer consolidation, turns a strong offline diarizer into an online one that recovers most of the online-to-offline error gap on the end-of-meeting track, and that this holds cross-domain under frozen calibration. We also show that the gate without consolidation does not achieve this. Second, we measure the system's behaviour rather than only its error rate, and find a coherent memory-like profile: fast one-shot recall, habituation of the novelty response, and recognition that survives long silences. Third, we reproduce on audio the visual SGM finding that consolidation must replay the full history, not a recent window, and we locate precisely where the residual error of the frozen backbone sits.

# 2 Background

## 2.1 Surprise-gated memory

The parent work computes a prediction-error signal over a frozen backbone and uses it as a write gate: a concept enters the episodic store only when surprise is high, and a periodic replay phase consolidates recent traces into a slow readout [1]. Two results from that work matter here. The surprise detector separates known from novel inputs at an AUROC of 0.966, well above the model's own verbalised confidence, and the consolidation is load-bearing: replaying only a recent window is worse than not replaying at all. Both transfer to the diarization setting, and we test both here.

The framing is the complementary-learning-systems account of memory [2], [3], in which a fast hippocampal store admits one-shot episodes and a slow neocortical store is consolidated by offline replay. The novelty signal that gates the fast store is the computational analogue of the orienting response [4], which habituates as a stimulus becomes familiar, and the retrieval of a full trace from a partial cue is pattern completion in the sense of [5]. These are the behaviours we measure in the section on memory-like descriptors.

## 2.2 Frozen backbone for diarization

The perception stack is Echo Dia, a fine-tuned DiariZen model [6] whose diarization component we treat as a fixed feature and clustering engine. DiariZen couples a WavLM self-supervised encoder [7] with a local Conformer end-to-end segmentation head and a Variational Bayes HMM (VBx) clusterer [8]; the pipeline decomposition follows the pyannote line of work [9]. Echo Dia shares the shared-latent-space philosophy of the Echo architecture [10], where a single encoder serves speaker identity and content. We use its segmentation head, its speaker embedder, and its VBx clusterer without modification, and no gradient reaches any of them.

## 2.3 Online diarization

The standard online formulation assigns each incoming segment to the nearest known speaker or opens a new one under a single distance threshold. This is our naive baseline B1. It is brittle: a single global threshold cannot be both permissive enough to re-identify a returning speaker and strict enough to avoid merging two similar ones, and it can never revise a premature commitment.

Prior online systems attack this from three directions. Speaker-tracing buffers carry a window of past frames into the end-to-end model so that permutations stay consistent across chunks [11]. Overlap-aware low-latency pipelines run local end-to-end segmentation and match incrementally against running centroids [12]. Graph-based online clustering generates labels from a similarity graph built over the stream so far [13]. All three keep a single accept-or-create decision rule and, with the partial exception of the buffered variants, commit irrevocably. The SGM formulation differs in two ways: it replaces the single threshold with a three-regime gate that can explicitly *defer*, and it adds a periodic consolidation phase that re-solves the whole assignment problem from the full buffer, which makes past decisions revisable.

## 3 Method

The system consumes the meeting as a stream of fixed chunks. For each chunk, the frozen backbone produces local speaker activity, including overlap, and one speaker embedding per active local speaker. Each embedding is projected into the backbone's LDA/PLDA space, where the surprise gate and the memory operate.

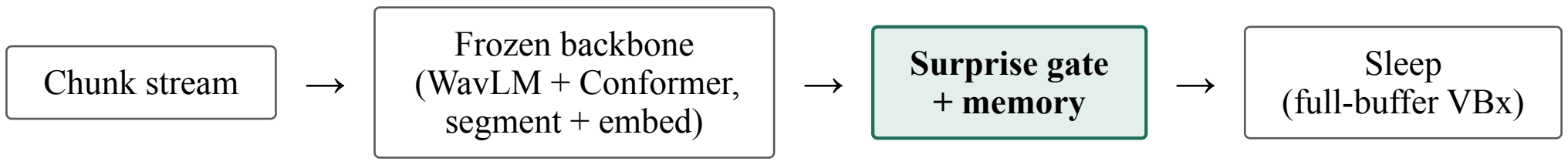

**Calibrated surprise.** The raw novelty of an embedding is its PLDA score against the nearest current prototype, negated so that higher means more novel. This raw score is z-scored against the inter-speaker score distribution estimated once on development data, in the spirit of adaptive score normalisation [14], so that the gate thresholds are expressed in standard deviations above the typical between-speaker score rather than in raw PLDA units. Computing this in the backbone's own PLDA space keeps the online metric aligned with the geometry the clusterer uses. On development data the known-versus-new detector reaches an AUROC of 0.866, below the visual detector's 0.966 because speaker embeddings from short far-field segments are markedly noisier than whole-image features.

**Three-regime gate.** Each embedding is routed by its surprise. Low surprise triggers an assertive assign to the nearest prototype with a running-mean update. Intermediate surprise defers the decision: the segment is held without emitting a label, to be resolved at the next consolidation. High surprise creates a new prototype, one-shot. Segments shorter than half a second of net speech, or flagged as overlapped, are forced to defer, because their embeddings are unreliable and must neither create nor move a prototype.

**Episodic memory.** Prototypes are non-parametric, written once and updated only by a running mean, never by gradient. The memory is strictly intra-meeting: created at the start, wiped at the end. A unit test verifies that no prototype and no stored trace survives the wipe, and that similarity queries return empty afterward.

**Sleep consolidation.** Every fixed interval of audio, the system replays the entire buffer accumulated since the start of the meeting and re-clusters it with VBx [8], merging prototypes that were split, re-estimating centroids, and resolving deferred segments. The choice of VBx as the sleep engine is deliberate. Plain agglomerative clustering over these embeddings over-segments badly, producing dozens of clusters for a handful of speakers, because single-embedding speaker purity is only about 0.73. The oracle's advantage comes not from embedding preprocessing but from the temporal HMM smoothing in VBx, so consolidation runs VBx itself over the buffer, which lifts cluster purity to roughly 0.79 to 0.87 and stays within the frozen pipeline.

The system emits two tracks. The causal-strict track fixes a label the moment it is emitted. The revisable track reflects the memory state after the last consolidation and is the meeting-minutes product.

## 4 Experimental setup

**Data.** We evaluate on two corpora with different acoustic and conversational character. AMI [15] is multi-party meeting speech; we use the single distant microphone (SDM) condition, which is the harder far-field case and the one that matches the meeting-minutes product. VoxConverse [16] is in-the-wild broadcast and interview audio with different speaker turn statistics and recording conditions. The test subsets are three AMI meetings and five VoxConverse sessions; the AMI development split is used for calibration only and never for evaluation.

**Calibration protocol.** All gate thresholds and surprise statistics are estimated once on the AMI development split and then frozen. They are applied unchanged to AMI test *and* to VoxConverse. No quantity is re-fitted per corpus, per meeting, or per speaker, so the VoxConverse results are a cross-domain transfer rather than a tuned result.

**Metric.** We report diarization error rate (DER), the sum of missed speech, false-alarm speech, and speaker confusion over total reference speech time. We score at collar zero with overlapped speech included, the strict setting used by recent challenge protocols [17]. This is deliberately unforgiving: it charges the system for every overlapped frame the frozen segmentation misses, which, as the section on the residual error shows, is where the residual error lives.

**Two output tracks.** *Causal-strict* fixes each label at emission time and never revises it, which is the correct metric for a live caption. *Revisable* reflects the memory state after the final consolidation, which is the correct metric for meeting minutes produced at the end of the session. Both are reported in Table 2; the revisable track is the intended product, and the causal track is reported for completeness rather than as a competitive result, for reasons the next section makes explicit.

**Hyperparameters.** The backbone is a DiariZen pipeline with a WavLM-large encoder and a Conformer end-to-end segmentation head operating on 16-second chunks, a wespeaker ResNet34 embedder, and VBx with Fa 0.07, Fb 0.8, LDA dimension 128 and AHC initialisation threshold 0.6. These are the backbone's own published settings and we change none of them. The surprise gate uses two thresholds on the z-scored novelty, 0.15 and 0.45, separating the assign, defer and create regimes; the naive baseline B1 uses a single cut at 0.30. Segments with under 0.5 s of net speech, or flagged as overlapped, are forced to defer. Prototypes are updated by a running mean over their assigned embeddings. The headline results use a sleep period of 60 seconds; the sweep reports 60, 120 and 300 seconds and an end-of-meeting-only variant.

**Conditions.** The comparisons use a fixed set of labelled conditions (Table 1).

Table 1: Condition labels used throughout. All share the same frozen backbone and the same frozen calibration.

| Label | Condition |
|---|---|
| B0 | Offline oracle: the full frozen pipeline with VBx over the entire meeting. Best achievable on this backbone. |
| B1 | Naive online: single-threshold nearest-prototype assignment, no deferral, no consolidation. Weakest online variant. |
| C1 | SGM-Diar: three-regime surprise gate with full-buffer sleep consolidation. |
| A1 | Ablation, no sleep: the gate without any consolidation phase. |
| A2 | Ablation, sliding-window sleep: consolidation replays only the last two minutes of buffer. |
| A3 | Ablation, no deferral: the gate keeps only its assign and create regimes, with full-buffer sleep. |
| A4 | Sweep of the sleep period, not a fixed condition. Reported separately. |
| C1+forget | Memory variant of C1 adding selective forgetting: prototype strength decays and prototypes not re-heard are pruned. Used only in the behavioural analysis. |

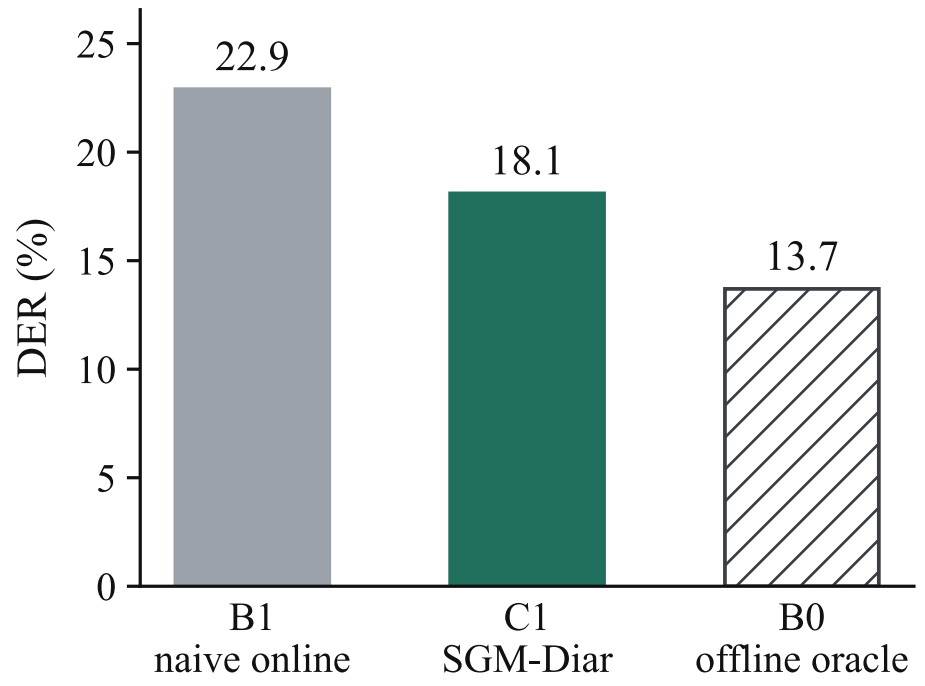

(a) AMI-SDM test

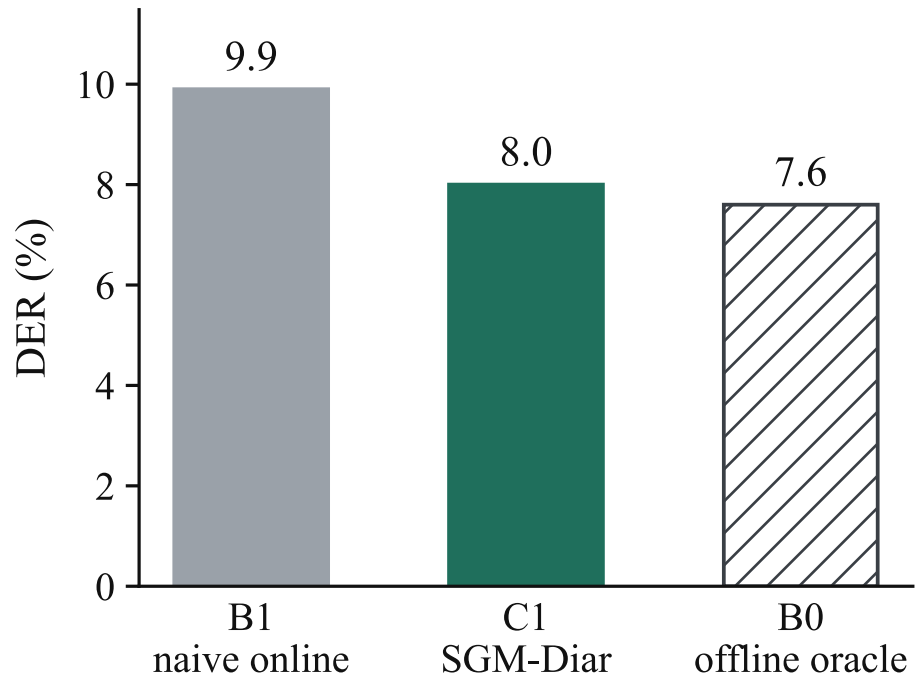

(b) VoxConverse test

Figure 1: Diarization error rate (collar 0, overlap scored) for the naive online baseline B1, SGM-Diar in revisable mode C1, and the offline oracle B0. Consolidation recovers 52% of the baseline-to-oracle gap on AMI-SDM and 82% on VoxConverse under cross-domain frozen calibration.

## 5 Accuracy: consolidation recovers the online-to-offline gap

Figure 1 shows the central measurement. In revisable mode it places the surprise-gated system between the naive online baseline and the offline oracle on both corpora, under the frozen cross-domain calibration described above.

The consolidation carries the result, and two ablations isolate its contribution (Table 2, Figure 2).

Replaying only the last two minutes of buffer (A2) is worse than full-buffer replay and worse even than no replay at all (A1), on both datasets. This reproduces on audio the central prediction of the visual SGM work, that a sliding replay window is actively harmful. The likely reason is that a recent window discards the early evidence that anchors a stable global assignment, so each consolidation re-solves a different, impoverished problem. We did not isolate this mechanism directly, and the magnitudes differ by a factor of three between corpora (2.7 points on AMI, 7.3 on VoxConverse), so we report the ordering as the result and the mechanism as a hypothesis. One observation cuts against a naive reading of it: A2 is markedly *better* than the full system on the causal track (39.2 against 58.2 on AMI). A window that re-solves a smaller problem also renumbers fewer identities, so the sliding window trades revisable quality for causal stability rather than being uniformly worse. Note also that the parent result concerned displacement in a parametric readout, whereas nothing here is parametric; the ordering transfers, the underlying mechanism cannot be the same one.

Two further readings of the table matter, and both cut against a simple story.

**The gate alone is not sufficient.** Without consolidation, the gate improves on the naive baseline on AMI by 2.1 points but *costs* 1.2 points on VoxConverse (A1 11.1 against B1 9.9). Deferred segments that are never resolved are simply lost, and on a corpus where the naive baseline is already close to the oracle that loss outweighs the gate's gains. The recovery we report is therefore a property of deferral *paired with* full-buffer consolidation, not of the surprise gate on its own.

Table 2: Diarization error rate (collar 0, overlap scored) by condition, on the AMI-SDM and VoxConverse test subsets. Lower is better. Revisable is the end-of-meeting product track; causal-strict fixes each label at emission. Calibration is frozen on AMI dev for every column.

| | Revisable | | Causal-strict | |
|---|---|---|---|---|
| Condition | AMI | Vox | AMI | Vox |
| B0 oracle offline | 13.7 | 7.6 | n/a | n/a |
| C1 full-buffer sleep | **18.1** | **8.0** | 58.2 | 47.8 |
| A1 no sleep | 20.8 | 11.1 | 20.8 | 11.1 |
| A2 sliding-window sleep | 23.5 | 18.4 | 39.2 | 44.9 |
| A3 no deferral | 17.1 | 7.7 | 62.8 | 46.1 |
| B1 naive online | 22.9 | 9.9 | 22.9 | 9.9 |

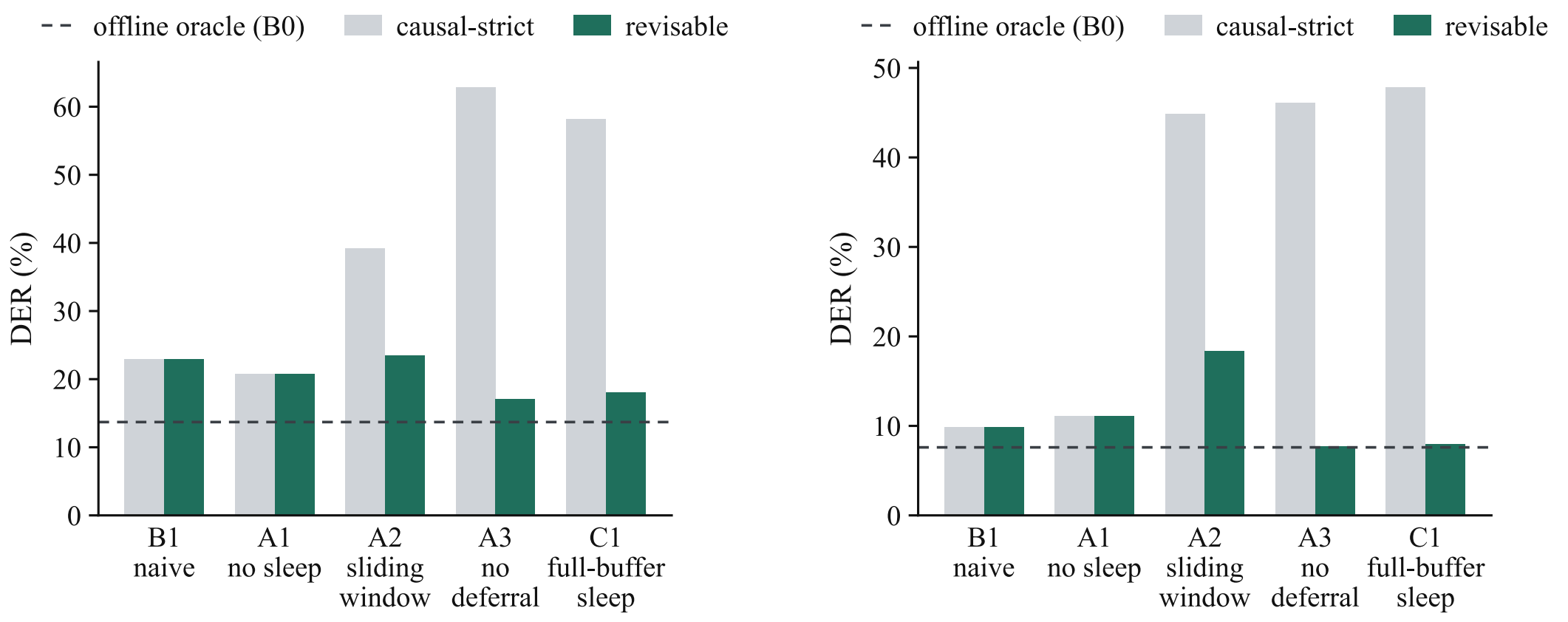

(a) AMI-SDM test (b) VoxConverse test

Figure 2: Diarization error rate by condition, causal versus revisable. The full-buffer sleep conditions approach the oracle line in revisable mode; the sliding-window ablation is the worst of the sleep conditions; causal error is inflated by cluster-identity renumbering across consolidations.

**Deferral is neutral for the product track.** Removing the deferral regime entirely (A3) leaves revisable error statistically indistinguishable from the full system, and nominally slightly better on both corpora (17.1 against 18.1 on AMI, 7.7 against 8.0 on VoxConverse). On three meetings and five sessions, single-seed, we do not read a one-point difference as real. The honest reading is that full-buffer VBx consolidation already resolves the ambiguous segments that deferral was designed to hold back, so on the revisable track deferral buys nothing. Its cost is visible on the causal track, where A3 is the worst condition of all (62.8 on AMI). We keep deferral in the headline system because it is what makes the causal track's held segments explicit rather than silently wrong, but we do not claim a revisable-error gain for it.

**The causal track is not competitive, and we say so.** Causal-strict error for the full system is 58.2 on AMI and 47.8 on VoxConverse, two to five times worse than the naive baseline it is meant to improve on. This is not a subtle degradation and we do not present it as one. The cause is identity churn rather than perception: every consolidation re-runs VBx over the whole buffer and is free to renumber clusters, so labels already emitted under the old numbering are scored as confusion even when the underlying partition improved. The sweep in Figure 3 isolates exactly this. A system that needs live captions would have to add a label-stability layer, for instance Hungarian matching of cluster identities between consecutive sleeps, which we did not implement in order to keep the causal semantics simple and honest. As it stands, the revisable track is the product and the causal track is a diagnostic.

Sweeping the sleep period leaves revisable error nearly flat while trading causal error against emission latency. A longer period renumbers cluster identities less often, which lowers causal error sharply, from 58.2 at a 60-second period to 25.9 when consolidation runs only at the end of the meeting, but defers labels for longer, with mean deferral latency rising above 900 seconds in the end-only setting. This is a quality-latency knob rather than a free parameter, and it confirms that the causal penalty is renumbering, not misperception.

## 6 Behaviour: memory-like descriptors of the gate

Error rate does not capture whether the system behaves like an episodic memory. We instrumented every gate decision against the reference speaker identities on the AMI test set and measured six descriptors chosen in advance (Table 3), each the operational counterpart of a property the complementary-learning-systems account predicts [2]. We use the cognitive terms as names for these measured quantities, not as claims about mechanism. A *turn* is a maximal run of speech attributed to one reference speaker with no intervening speech from another, and averages about five seconds on AMI, which is how 1.8 turns becomes roughly nine seconds below. A *gap* is the silence between two consecutive turns of the same speaker, and the gap re-ID descriptor aggregates returns after more than fifteen seconds. *One-shot recall* is the fraction of a speaker's turns after their first appearance that are routed to the prototype created on that first appearance. *False novelty* is the fraction of gate decisions that fire NEW on a speaker already in memory.

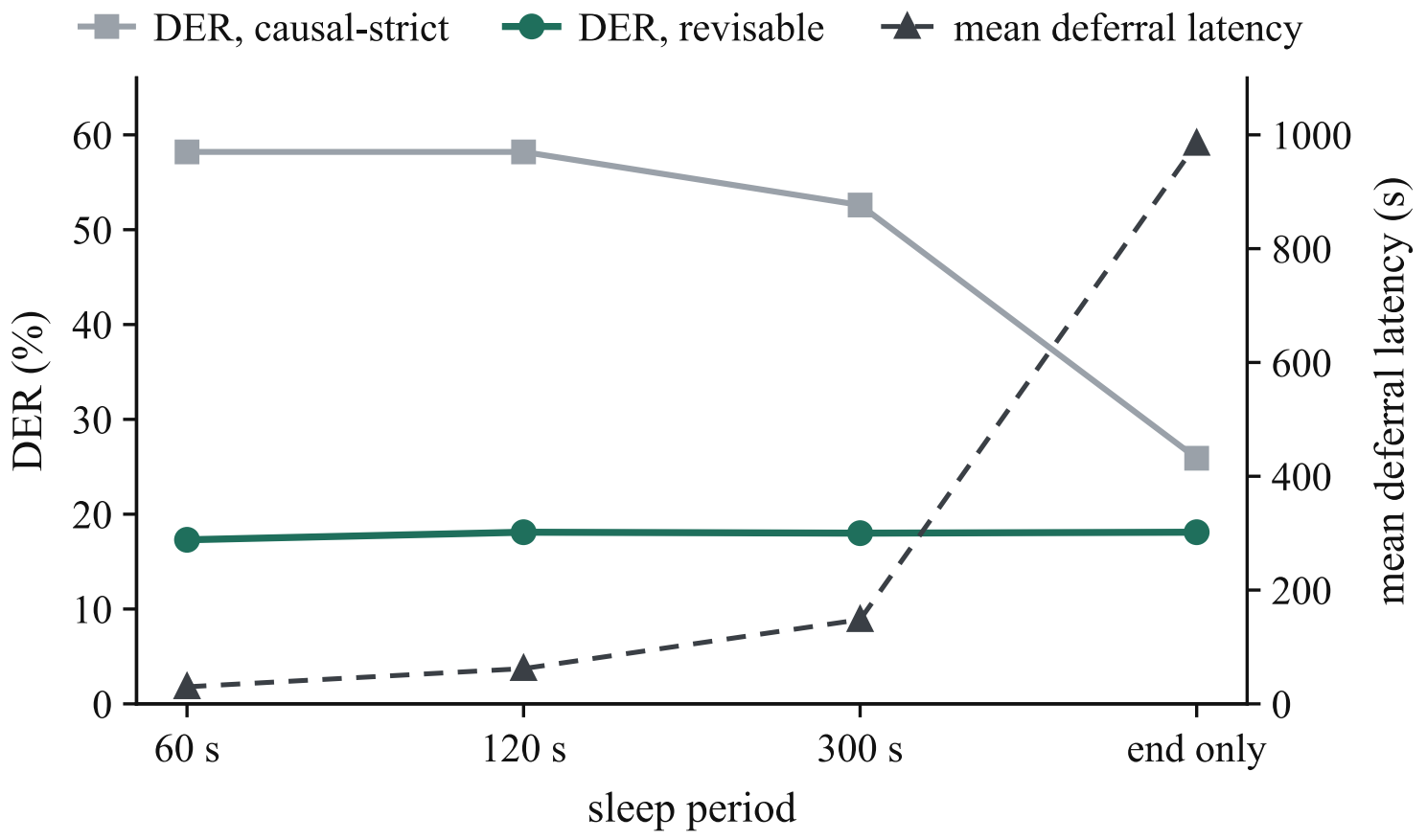

Figure 3: Sleep-period sweep on AMI-SDM. Revisable error is flat across the period; causal error falls as the period lengthens while deferred-segment latency rises.

Figure 4 summarises these. The mean surprise over a meeting falls as speakers become known: it starts high and habituates as speakers become known, and it re-spikes when a new speaker enters. Every speaker's first-contact window scores above their later turns.

The most suggestive result is re-identification across silence (Figure 5). Recognition of a returning speaker stays essentially flat, between roughly 0.76 and 0.87, from fifteen seconds of silence out to over two minutes, with a correlation between recognition and gap length, computed per return over all events rather than over the bins plotted below, of about negative 0.02. Computed on so few long-gap events this statistic is dominated by the short-gap bin and we do not lean on it. The counts thin out quickly with gap length: of the returns after more than fifteen seconds of silence there are about 149, of which 28 follow a gap longer than two minutes, so the flatness at the long end rests on few events and we do not attach a confidence interval to it. Recognition independent of how long a voice was silent is what we would expect of pattern completion, though on these counts we treat it as consistent with that reading rather than as evidence for it. The C1+forget variant, which decays and prunes prototypes that are not re-heard, shows the complementary signature: its re-identification declines only at the longest gaps, which is what a memory that forgets should do.

Table 3: Behavioural descriptors measured against reference speaker identities on the AMI-SDM test set. These are our own operational definitions, not a standard benchmark.

| Descriptor | Value | Reading |
|---|---|---|
| One-shot recall rate | 0.857 | turns routed correctly after first contact |
| First-recall latency | 1.8 turns | about nine seconds to lock identity |
| Surprise early to late | 0.124 to 0.061 | novelty response habituates |
| Gap re-ID (>15 s silence) | 0.81 | recognition survives long silence |
| Novelty gap (novel − known) | 0.116 | clean new-versus-familiar separation |
| False-novelty rate | 0.001 | fires NEW on an already-known speaker in about one gate decision in a thousand |

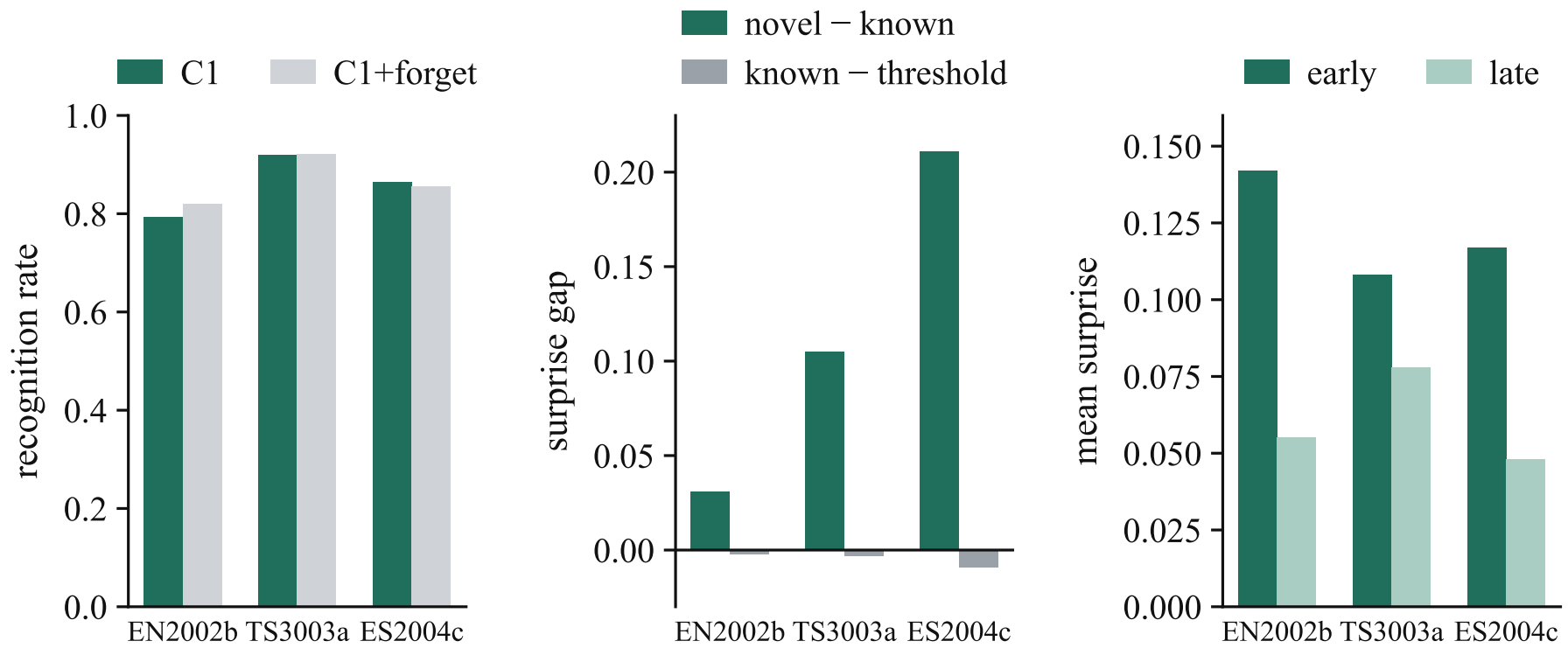

(a) One-shot recall (b) Novelty signal (c) Habituation

Figure 4: One-shot recall, novelty signal, and habituation on AMI-SDM test. Recognition after first contact is high (a); the novelty gap between new and known speakers is positive, while known speakers sit below the NEW threshold by a small margin (b); early-meeting surprise exceeds late-meeting surprise (c).

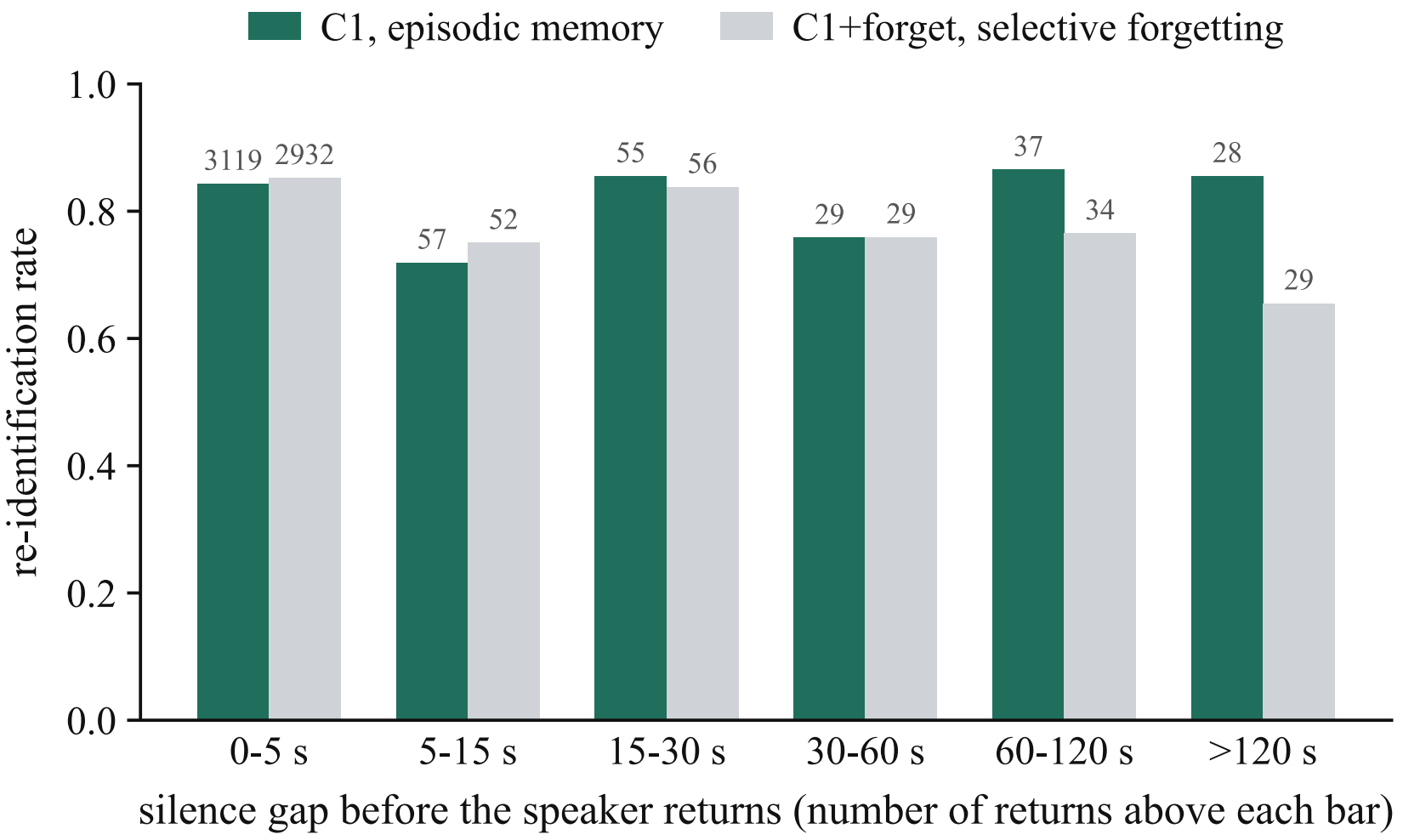

Figure 5: Re-identification rate as a function of intervening silence gap. Recognition is independent of gap length for the base episodic memory, consistent with pattern completion; the C1+forget variant declines at the longest gaps.

## 7 Comparison with published online systems

The ablations above isolate the contribution of the memory against our own baselines on a fixed backbone. They do not say where the system sits relative to the online diarization literature. A like-for-like comparison is not available, for reasons we state before giving the numbers, but one published system is close enough on protocol to be worth putting side by side.

The comparison is with diart [12], which is evaluated with pyannote.metrics at collar zero with overlapped speech scored, the same strict setting we use. Two differences matter and neither is in our disfavour by accident.

**The AMI condition differs.** diart reports AMI Mix-Headset, a mixed close-talking condition. We report AMI-SDM, the single distant microphone, which is the harder far-field case. The two AMI columns below are therefore *not* comparable and we do not treat the AMI difference as a result. VoxConverse is the same corpus and the same protocol in both, so that column is comparable.

**The backbones differ, and this is the larger effect.** diart is built on the 2021 pyannote segmentation and embedding models; our frozen backbone is a 2025 DiariZen model [6] whose offline oracle is correspondingly stronger. Most of the absolute distance between the two rows is attributable to perception rather than to memory. The clearest evidence is in our own naive baseline: B1 on VoxConverse already reaches 9.9 without any of the machinery this note describes, which is better than diart's online result. What this note contributes is the improvement *over that baseline on the same backbone*, which is the 52 to 82% gap recovery reported above, not the distance to diart.

Table 4: Online systems, DER at collar 0 with overlapped speech scored (pyannote.metrics) in both cases. The AMI columns use different conditions and are not comparable: diart reports Mix-Headset, we report the harder SDM. The VoxConverse column is like-for-like on corpus and protocol, but not on subset or backbone. Offline toplines are given only as each system's own ceiling.

| System | Latency | AMI | Vox |
|---|---|---|---|
| *diart, pyannote backbone, AMI Mix-Headset* | | | |
| diart online | 5 s | 27.5 | 16.8 |
| diart online | 1 s | 30.4 | 20.1 |
| its offline topline (VBx + overlap reseg.) | offline | 19.9 | 11.1 |
| *This work, DiariZen backbone, AMI-SDM* | | | |
| B1 naive online | chunk | 22.9 | 9.9 |
| C1 SGM-Diar revisable | end of mtg. | **18.1** | **8.0** |
| its offline oracle | offline | 13.7 | 7.6 |

One difference runs the other way. diart optimises three pipeline hyperparameters separately on each corpus's development set, whereas every threshold in our system is frozen on AMI dev and transferred to VoxConverse untouched. On that column the comparison is therefore conservative in our favour, which we note for completeness rather than as a claim.

Two further caveats bound Table 4. Our VoxConverse figure is computed on five sessions whereas diart reports the full test set, and several versions of that test set circulate, so the column is like-for-like on corpus and scoring protocol but not on subset. The three AMI meetings and five VoxConverse sessions were fixed before any of these experiments were run and were not selected on performance.

The honest reading of Table 4 is narrow. On VoxConverse, where the comparison is valid on data and protocol, SGM-Diar reaches 8.0 against diart's 16.8 at five seconds of latency, but our own untuned baseline already reaches 9.9 on the same backbone, so the backbone accounts for most of that distance and the memory for the remainder. On AMI the conditions differ and we draw no conclusion. We report Table 4 to situate the work, not to claim a ranking, and a genuine like-for-like comparison would require running both systems on the same frozen backbone and the same audio condition, which we have not done.

Our latency is also not diart's latency. diart emits a label within one to five seconds of the audio arriving. Our revisable track is an end-of-meeting product: labels are final only after the last consolidation. The causal-strict track is the one that would be compared against diart's operating point, and as reported above it is not competitive. These systems answer different product questions.

## 8 Positioning

SGM-Diar sits between two literatures that each hold part of it.

Against **online diarization**, the difference is the decision rule, not the backbone. Speaker-tracing buffers [11] and incremental centroid matching [12] both carry context forward to keep labels consistent, and graph-based online clustering [13] accumulates a similarity structure over the stream; all three still make a binary accept-or-create choice per segment and propagate it forward. SGM-Diar adds a third option, deferral, for segments the embedding cannot resolve, which is precisely the short and overlapped ones, and a consolidation that re-solves the assignment from the full buffer rather than extending it incrementally. The price is that a label may change before it is final, which is why we report the causal and revisable tracks separately rather than reporting only the favourable one.

Against the **memory-and-consolidation** literature, the difference is that the consolidation engine is the frozen pipeline's own clusterer. Replay-based continual learning and its neuroscientific framing [2], [18] consolidate into a parametric slow store by gradient. Here no gradient exists at run time: the slow store is a set of non-parametric prototypes and the replay step is VBx over the buffer. This buys the privacy property below and rules out representation drift by construction, at the cost that the memory cannot outperform the clusterer it delegates to, a boundary we make explicit in the next section.

The specific combination we report, a calibrated three-regime novelty gate over a frozen speaker embedder with full-buffer consolidation and strict intra-session memory, does not appear as a whole in prior work, though each individual piece does.

## 9 Where the residual error lives

The offline oracle is the ceiling this system is built against, not a competitor: it sees the whole meeting before deciding, which an online system by definition cannot. The question worth asking is where the remaining error lives, because that determines what a better memory could and could not fix. It turns out that almost none of it is the memory's to win. Decomposing the oracle's 13.7 points on AMI, clustering confusion is already small at roughly 2 to 2.5 points, missed overlapped speech accounts for about 6.8, and the remaining 4.5 or so is voice-activity false alarm. Confusion is the only term the clustering or the memory controls, and it is the smallest. The other two are properties of the frozen segmentation.

Feeding the true instantaneous speaker count to the reconstruction drops AMI error from 13.7 to 6.6. That intervention recovers 7.1 points, slightly more than the missed-overlap term alone, because knowing the exact number of active speakers at each instant also tells the system when nobody is speaking and so removes voice-activity false alarm as well. The experiment therefore bounds the combined overlap-plus-VAD prize rather than isolating overlap, and we did not run an overlap-only oracle to separate them. What it does establish is that the prize lies in the segmentation, not in the clustering or in the memory.

**The residual obstacle: overlapped speech.** The second speaker in an overlapped frame has a mixture embedding of two voices, so neither its cluster activation nor a re-extracted embedding identifies it cleanly, and including overlapped frames in the clustering pollutes the centroids. Across eleven post-hoc and light fine-tune attempts, none improved offline error: assigning the correct second speaker reaches only about 15 to 17% precision, so added false alarms exceed recovered misses. Approach envisaged: full retraining of the WavLM and Conformer representation, or dedicated target-speaker extraction, which is out of scope for a frozen-backbone system.

Two further cognitive mechanisms were tested as memory ablations. Selective forgetting (C1+forget, Figure 5) improves the causal track and the speaker-count estimate everywhere, but improves revisable error only where over-segmentation exists, because the full-buffer sleep re-instates forgotten traces. Prioritised replay, which repeats surprising or recent traces more often in the consolidation, over-segments, because replicating a trace makes one speaker look like several to the clusterer. With a full-buffer oracle clusterer at sleep, the quality output is dominated by that clusterer, so cognitive memory dynamics change behaviour more than they change error rate. This is a boundary of the approach, stated rather than hidden.

## 10 Privacy by construction

Speaker memory exists only within a meeting and is wiped at its end, verified by a unit test that finds zero prototypes, zero stored traces, and empty similarity queries afterward. The run-time system performs no parametric learning: no gradient, no fine-tuning, no test-time adaptation. The only fitted quantities are development-set surprise statistics, computed once and frozen. No speaker representation persists across sessions.

## 11 Limitations

Several limitations bound the claims.

**Statistical power.** The test subsets are small, three meetings for AMI and five for VoxConverse, and runs are single-seed with no confidence intervals. The load-bearing result is therefore the *ordering* of conditions, which is consistent across two corpora with different acoustics, and not the absolute error rates. The margins we treat as real are the large ones: full-buffer sleep against no sleep, and against the sliding window. We would not defend a difference of a few tenths of a point.

**No like-for-like external baseline.** Our ablations compare against our own naive online baseline and our own offline oracle on the same frozen backbone, which isolates the contribution of the gate and the consolidation. Table 4 adds one published online system evaluated under the same scoring protocol, but it differs from us in backbone and, on AMI, in recording condition, so it situates the work rather than ranking it. We did not re-run any published system on our backbone, and this note therefore establishes an internal effect, not a place on a leaderboard.

**Causal track.** The causal-strict numbers are poor in absolute terms (58.2 on AMI, 47.8 on VoxConverse, against 22.9 and 9.9 for the naive baseline), and we report them rather than omitting them. Two distinct causes are involved: single-embedding speaker purity near 0.73, which bounds any single-shot decision, and cluster-identity renumbering across consolidations, which the period sweep shows dominates. The revisable track is the intended product and is unaffected, but any caption-style application would need identity tracking across sleeps before these numbers mean anything.

**Behavioural descriptors.** The cognitive measurements depend on how hypothesis segments are aligned to reference speakers and on the thresholds that define a "turn" and a "gap". They are internally consistent and were fixed before measurement, but they are our operational definitions rather than a standard benchmark, and the absolute values should not be compared across papers.

**Scope.** Overlap is handled only by the frozen backbone, and the episodic memory does not model it. The comparison is not like-for-like against a from-scratch online system, since the backbone is pretrained and fine-tuned.

## 12 Conclusion

SGM-Diar shows that the Surprise-Gated Memory recipe, frozen perception with a calibrated novelty gate over a one-shot episodic store consolidated by full replay, transfers from continual visual learning to live speaker diarization and recovers most of the online-to-offline error gap under frozen cross-domain calibration. Measured as behaviour rather than error rate, it recalls a speaker after one exposure, habituates to familiar voices, and re-identifies a returning speaker regardless of how long they were silent, a profile consistent with the complementary-learning-systems account that motivated the design. The remaining gap to the offline ceiling is overlapped-speech miss living in the frozen representation, not anything the memory controls, which bounds what any online memory over this backbone could achieve. The contribution is a way to run diarization live that behaves like a memory, and a precise account of where the hard error remains.